\input harvmac

\input graphicx
\input color

\def\Title#1#2{\rightline{#1}\ifx\answ\bigans\nopagenumbers\pageno0\vskip1in
\else\pageno1\vskip.8in\fi \centerline{\titlefont #2}\vskip .5in}

\font\titlerm=cmr10 scaled\magstep3 \font\titlerms=cmr7 scaled\magstep3
\font\titlermss=cmr5 scaled\magstep3 \font\titlei=cmmi10 scaled\magstep3
\font\titleis=cmmi7 scaled\magstep3 \font\titleiss=cmmi5 scaled\magstep3
\font\titlesy=cmsy10 scaled\magstep3 \font\titlesys=cmsy7 scaled\magstep3
\font\titlesyss=cmsy5 scaled\magstep3 \font\titleit=cmti10 scaled\magstep3
%
%
\ifx\includegraphics\UnDeFiNeD\message{(NO graphicx.tex, FIGURES WILL BE IGNORED)}
\def\figin#1{\vskip2in}
\else\message{(FIGURES WILL BE INCLUDED)}\def\figin#1{#1}
\fi
\def\Fig#1{Fig.~\the\figno\xdef#1{Fig.~\the\figno}\global\advance\figno
 by1}
%
%
%
%
\def\Ifig#1#2#3#4{
\goodbreak\midinsert
\figin{\centerline{
\includegraphics[width=#4truein]{#3}}}
\narrower\narrower\noindent{\footnotefont
{\bf #1:}  #2\par}
\endinsert
}

\font\ticp=cmcsc10

\def \purge#1 {\textcolor{magenta}{#1}}
\def \new#1 {\textcolor{blue}{#1}}
\def\comment#1{}

\def\\{\cr}
\def\text#1{{\rm #1}}
\def\frac#1#2{{#1\over#2}}

\def\subsubsec#1{\noindent{\undertext {#1}}}
\def\undertext#1{$\underline{\smash{\hbox{#1}}}$}
\def\hf{{1\over 2}} 
\def\calo{{\cal O}}

\def\roughly#1{\mathrel{\raise.3ex\hbox{$#1$\kern-.75em\lower1ex\hbox{$\sim$}}}}
\font\bbbi=msbm10 
\def\mathbb#1{\hbox{\bbbi #1}}

\def\mthsu{\mathsurround=0pt  }
\def\leftrightarrowfill{$\mthsu \mathord\leftarrow\mkern-6mu\cleaders
  \hbox{$\mkern-2mu \mathord- \mkern-2mu$}\hfill
  \mkern-6mu\mathord\rightarrow$}
\def\overleftrightarrow#1{\vbox{\ialign{##\crcr\leftrightarrowfill\crcr\noalign{\kern-1pt\nointerlineskip}$\hfil\displaystyle{#1}\hfil$\crcr}}}
\overfullrule=0pt

%
%
\lref\Hawkrad{
  S.~W.~Hawking,
  ``Particle Creation By Black Holes,''
  Commun.\ Math.\ Phys.\  {\bf 43}, 199 (1975)
  [Erratum-ibid.\  {\bf 46}, 206 (1976)].
}
\lref\NVU{S.~B.~Giddings,
``Nonviolent unitarization: basic postulates to soft quantum structure of black holes,''
JHEP{\bf 12}, 047 (2017)
doi:10.1007/JHEP12(2017)047
[arXiv:1701.08765 [hep-th]].}
\lref\SGStBo{S.~B.~Giddings,
``Hawking radiation, the Stefan-Boltzmann law, and unitarization,''
Phys. Lett. B {\bf 754}, 39-42 (2016)
doi:10.1016/j.physletb.2015.12.076
[arXiv:1511.08221 [hep-th]].}
\lref\SGTrieste{
S.~B.~Giddings,
``Quantum mechanics of black holes,''
[arXiv:hep-th/9412138 [hep-th]].}
\lref\AMPS{
  A.~Almheiri, D.~Marolf, J.~Polchinski and J.~Sully,
  ``Black Holes: Complementarity or Firewalls?,''
  JHEP {\bf 1302}, 062 (2013).
  [arXiv:1207.3123 [hep-th]].
}
\lref\AMPSS{
A.~Almheiri, D.~Marolf, J.~Polchinski, D.~Stanford and J.~Sully,
``An Apologia for Firewalls,''
JHEP {\bf 09}, 018 (2013)
doi:10.1007/JHEP09(2013)018
[arXiv:1304.6483 [hep-th]].}
\lref\PaWi{
M.~K.~Parikh and F.~Wilczek,
``Hawking radiation as tunneling,''
Phys. Rev. Lett. {\bf 85}, 5042-5045 (2000)
doi:10.1103/PhysRevLett.85.5042
[arXiv:hep-th/9907001 [hep-th]].}
\lref\CGHS{C.~G.~Callan, Jr., S.~B.~Giddings, J.~A.~Harvey and A.~Strominger,
``Evanescent black holes,''
Phys. Rev. D {\bf 45}, no.4, 1005 (1992)
doi:10.1103/PhysRevD.45.R1005
[arXiv:hep-th/9111056 [hep-th]].}
\lref\UnruNotes{
W.~Unruh,
``Notes on black hole evaporation,''
Phys. Rev. D {\bf 14} (1976), 870
doi:10.1103/PhysRevD.14.870.}
\lref\BHQIUE{
  S.~B.~Giddings,
  ``Black holes, quantum information, and unitary evolution,''
  Phys.\ Rev.\ D {\bf 85}, 124063 (2012).
[arXiv:1201.1037 [hep-th]].
}
\lref\NLvC{
  S.~B.~Giddings,
  ``Nonlocality versus complementarity: A Conservative approach to the information problem,''
Class.\ Quant.\ Grav.\  {\bf 28}, 025002 (2011).
[arXiv:0911.3395 [hep-th]].
}
\lref\LPSTU{
  D.~A.~Lowe, J.~Polchinski, L.~Susskind, L.~Thorlacius and J.~Uglum,
  ``Black hole complementarity versus locality,''
Phys.\ Rev.\ D {\bf 52}, 6997 (1995).
[hep-th/9506138].
}
\lref\QBHB{
  S.~B.~Giddings,
  ``Quantization in black hole backgrounds,''
Phys.\ Rev.\ D {\bf 76}, 064027 (2007).
[hep-th/0703116 [HEP-TH]].
}
\lref\HawkUnc{
  S.~W.~Hawking,
  ``Breakdown of Predictability in Gravitational Collapse,''
Phys.\ Rev.\ D {\bf 14}, 2460 (1976).
}
\lref\Boul{D.~G.~Boulware,
``Quantum Field Theory in Schwarzschild and Rindler Spaces,''
Phys. Rev. D {\bf 11}, 1404 (1975)
doi:10.1103/PhysRevD.11.1404.}
\lref\Kay{
B.~S.~Kay,
``Linear Spin 0 Quantum Fields in External Gravitational and Scalar Fields. 1. A One Particle Structure for the Stationary Case,''
Commun. Math. Phys. {\bf 62}, 55-70 (1978)
doi:10.1007/BF01940330.}
\lref\ADM{R.~L.~Arnowitt, S.~Deser and C.~W.~Misner,
``The Dynamics of general relativity,''
Gen. Rel. Grav. {\bf 40}, 1997-2027 (2008)
doi:10.1007/s10714-008-0661-1
[arXiv:gr-qc/0405109 [gr-qc]].}
\lref\More{
C.~Moreno,
``Spaces of Positive and Negative Frequency Solutions of Field Equations in Curved Space-Times. 1. The Klein-Gordon Equation in Stationary Space-Times,''
J. Math. Phys. {\bf 18}, 2153-2161 (1977)
doi:10.1063/1.523197.}
\lref\ToVa{
C.~G.~Torre and M.~Varadarajan,
``Functional evolution of free quantum fields,''
Class. Quant. Grav. {\bf 16}, 2651-2668 (1999)
doi:10.1088/0264-9381/16/8/306
[arXiv:hep-th/9811222 [hep-th]].}
\lref\Cortetal{J.~Cortez, G.~A.~Mena Marugan, J.~Olmedo and J.~M.~Velhinho,
``A uniqueness criterion for the Fock quantization of scalar fields with time dependent mass,''
Class. Quant. Grav. {\bf 28}, 172001 (2011)
doi:10.1088/0264-9381/28/17/172001
[arXiv:1106.5000 [gr-qc]];
J.~Cortez, L.~Fonseca, D.~Martín-de Blas and G.~A.~Mena Marugán,
``Uniqueness of the Fock quantization of scalar fields under mode preserving canonical transformations varying in time,''
Phys. Rev. D {\bf 87}, no.4, 044013 (2013)
doi:10.1103/PhysRevD.87.044013
[arXiv:1212.3947 [gr-qc]], and references therein.}
\lref\AgAs{
I.~Agullo and A.~Ashtekar,
``Unitarity and ultraviolet regularity in cosmology,''
Phys. Rev. D {\bf 91}, no.12, 124010 (2015)\break
doi:10.1103/PhysRevD.91.124010
[arXiv:1503.03407 [gr-qc]].}
\lref\MuOe{A.~Much and R.~Oeckl,
``Self-Adjointness in Klein-Gordon Theory on Globally Hyperbolic Spacetimes,''
[arXiv:1804.07782 [math-ph]].}
\lref\BHIUN{
S.~B.~Giddings,
``Black hole information, unitarity, and nonlocality,''
Phys. Rev. D {\bf 74}, 106005 (2006)
doi:10.1103/PhysRevD.74.106005
[arXiv:hep-th/0605196 [hep-th]].}
\lref\GiNe{
S.~B.~Giddings and W.~M.~Nelson,
``Quantum emission from two-dimensional black holes,''
Phys. Rev. D {\bf 46}, 2486-2496 (1992)
doi:10.1103/PhysRevD.46.2486
[arXiv:hep-th/9204072 [hep-th]].}
\lref\BPZ{S.~L.~Braunstein, S.~Pirandola and K.~Zyczkowski,
``Better Late than Never: Information Retrieval from Black Holes,''
Phys. Rev. Lett. {\bf 110}, no.10, 101301 (2013)
doi:10.1103/PhysRevLett.110.101301
[arXiv:0907.1190 [quant-ph]].}
\lref\DLP{
R.~Dey, S.~Liberati and D.~Pranzetti,
``The black hole quantum atmosphere,''
Phys. Lett. B {\bf 774}, 308-316 (2017)
doi:10.1016/j.physletb.2017.09.076
[arXiv:1701.06161 [gr-qc]].}
\lref\Mathur{
  S.~D.~Mathur,
  ``The Information paradox: A Pedagogical introduction,''
Class.\ Quant.\ Grav.\  {\bf 26}, 224001 (2009).
[arXiv:0909.1038 [hep-th]].
}
\lref\DoGione{
  W.~Donnelly and S.~B.~Giddings,
  ``Diffeomorphism-invariant observables and their nonlocal algebra,''
Phys.\ Rev.\ D {\bf 93}, no. 2, 024030 (2016), Erratum: [Phys.\ Rev.\ D {\bf 94}, no. 2, 029903 (2016)].
[arXiv:1507.07921 [hep-th]].
}
\lref\MaPo{
K.~Martel and E.~Poisson,
``Regular coordinate systems for Schwarzschild and other spherical space-times,''
Am. J. Phys. {\bf 69}, 476-480 (2001)
doi:10.1119/1.1336836
[arXiv:gr-qc/0001069 [gr-qc]].}
\lref\Unru{W.~Unruh,
``Origin of the Particles in Black Hole Evaporation,''
Phys. Rev. D {\bf 15} (1977), 365-369
doi:10.1103/PhysRevD.15.365.}
\lref\Full{
S.~Fulling,
``Radiation and Vacuum Polarization Near a Black Hole,''
Phys. Rev. D {\bf 15} (1977), 2411-2414
doi:10.1103/PhysRevD.15.2411.}
\lref\JacoIntro{
T.~Jacobson,
``Introduction to quantum fields in curved space-time and the Hawking effect,''
doi:10.1007/0-387-24992-3\_2
[arXiv:gr-qc/0308048 [gr-qc]].}
\lref\SGmodels{
S.~B.~Giddings,
``Models for unitary black hole disintegration,''
Phys. Rev. D {\bf 85}, 044038 (2012)
doi:10.1103/PhysRevD.85.044038
[arXiv:1108.2015 [hep-th]]}
\lref\NVearly{
S.~B.~Giddings and Y.~Shi,
``Quantum information transfer and models for black hole mechanics,''
Phys. Rev. D {\bf 87}, no.6, 064031 (2013)
doi:10.1103/PhysRevD.87.064031
[arXiv:1205.4732 [hep-th]]; 
S.~B.~Giddings,
``Nonviolent nonlocality,''
Phys. Rev. D {\bf 88}, 064023 (2013)
doi:10.1103/PhysRevD.88.064023
[arXiv:1211.7070 [hep-th]]; 
``Nonviolent information transfer from black holes: A field theory parametrization,''
Phys. Rev. D {\bf 88}, no.2, 024018 (2013)
doi:10.1103/PhysRevD.88.024018
[arXiv:1302.2613 [hep-th]]; ``Modulated Hawking radiation and a nonviolent channel for information release,''
Phys. Lett. B {\bf 738}, 92-96 (2014)
doi:10.1016/j.physletb.2014.08.070
[arXiv:1401.5804 [hep-th]].}
\lref\WittBH{
E.~Witten,
``On string theory and black holes,''
Phys. Rev. D {\bf 44}, 314-324 (1991)
doi:10.1103/PhysRevD.44.314.}
\lref\JacoC{
T.~Jacobson,
``Black hole radiation in the presence of a short distance cutoff,''
Phys. Rev. D {\bf 48} (1993), 728-741
doi:10.1103/PhysRevD.48.728
[arXiv:hep-th/9303103 [hep-th]].}
\lref\Bard{
J.~M.~Bardeen,
``Black hole evaporation without an event horizon,''
[arXiv:1406.4098 [gr-qc]].}
\lref\JacoRev{
T.~Jacobson,
``Black holes and Hawking radiation in spacetime and its analogues,''
Lect. Notes Phys. {\bf 870}, 1-29 (2013)
doi:10.1007/978-3-319-00266-8\_1
[arXiv:1212.6821 [gr-qc]].}
\lref\MeWeone{K.~Melnikov and M.~Weinstein,
``A Canonical Hamiltonian derivation of Hawking radiation,''
[arXiv:hep-th/0109201 [hep-th]].}
\lref\MeWetwo{K.~Melnikov and M.~Weinstein,
``On unitary evolution of a massless scalar field in a Schwarzschild background: Hawking radiation and the information paradox,''
Int. J. Mod. Phys. D {\bf 13}, 1595-1636 (2004)
doi:10.1142/S0218271804005249
[arXiv:hep-th/0205223 [hep-th]].}
\lref\DFU{
P.~Davies, S.~Fulling and W.~Unruh,
``Energy Momentum Tensor Near an Evaporating Black Hole,''
Phys. Rev. D {\bf 13} (1976), 2720-2723
doi:10.1103/PhysRevD.13.2720.}
\lref\ChFu{
S.~Christensen and S.~Fulling,
``Trace Anomalies and the Hawking Effect,''
Phys. Rev. D {\bf 15} (1977), 2088-2104
doi:10.1103/PhysRevD.15.2088.}
\lref\LQGST{
  S.~B.~Giddings,
 ``Locality in quantum gravity and string theory,''
Phys.\ Rev.\ D {\bf 74}, 106006 (2006).
[hep-th/0604072].
}
\lref\BHS{
S.~Barman, G.~M.~Hossain and C.~Singha,
``Exact derivation of the Hawking effect in canonical formulation,''
Phys. Rev. D {\bf 97} (2018) no.2, 025016
doi:10.1103/PhysRevD.97. 025016
[arXiv:1707.03614 [gr-qc]].}
\lref\HoSi{G.~M.~Hossain and C.~Singha,
``New coordinates for a simpler canonical derivation of the Hawking effect,''
Eur. Phys. J. C {\bf 80} (2020) no.2, 82
doi:10.1140/epjc/s10052-020-7660-0
[arXiv:1902.04781 [gr-qc]].}
\Title{
\vbox{\baselineskip12pt  
}}
{\vbox{\centerline{Schr\"odinger evolution of the Hawking state} 
}}

\centerline{{\ticp 
Steven B. Giddings\footnote{$^\ast$}{Email address: giddings@ucsb.edu}
} }
\centerline{\sl Department of Physics}
\centerline{\sl University of California}
\centerline{\sl Santa Barbara, CA 93106}
\vskip.20in
\centerline{\bf Abstract} A Schr\"odinger-picture description of the  evolving quantum state of Hawking radiation is given, based on an ADM decomposition using time slicings that smoothly cross the horizon.  This treatment avoids requiring a role for trans-planckian excitations, which can be viewed as artifacts of Hawking's original calculation, and also supports arguments that radiation from black holes is produced in a ``quantum atmosphere" with thickness comparable to the horizon size, rather than microscopically far from it.  Particularly explicit formulas are given for the two-dimensional analog of the Schwarzschild geometry.  This analysis is expected to generalize to other black holes, and  to cosmology.  The resulting quantum evolution also provides important background for investigating corrections to the Hawking process, as are necessary for restoring unitary evolution of black hole decay.  

\vskip.3in
\Date{}

\newsec{Introduction}

Hawking's discovery\refs{\Hawkrad} that black holes radiate has had  continuing profound effect on the study of quantum gravity, in large part through its challenge to unitarity of quantum evolution.  Its central role implies the importance both of fully understanding the calculation, and of possible modifications to it which can restore unitarity.  

In particular, Hawking's original calculation exhibited certain pathologies, which have continued to be discussed and debated both for black holes and in the analogous treatment of cosmological production of fluctuations: a role for ultrahigh-energy excitations, in principle far beyond the Planck scale, as seen by observers falling through the BH horizon.\foot{For one discussion of this problem, with connections to the present analysis, see \JacoIntro.}

The role of these modes is related to the S-matrix form of Hawking's calculation, in which he considers specific outgoing modes and traces them back to an origin near the horizon.  In contrast, it has seemed  desirable to have a description of the dynamical evolution of the state, to have a more explicit description of the emergence of excitations from the near-horizon region.  Specifically, we might seek a Schr\"odinger picture description of the time-dependent evolution of the state near the black hole. One obstacle here is that the usual Schwarzschild time becomes pathological near the horizon, again leading one to consider ultrahigh energy modes.  But, other time slicings of the geometry exist, with better behavior near the horizon, and this suggests a regular dynamical description can be given by following evolution on such a slicing.

Having an improved description of the evolving quantum state of a black hole and its surroundings is also important because we seek to understand modifications to Hawking's evolution, which restore unitary evolution.  These are expected to take the form of corrections that transfer information (or entanglement) from the internal state of the black hole to the outgoing radiation.  If these are small corrections, in an appropriate sense, to the Hawking evolution, better understanding the background provided by the latter is a key first step to describing their effect\refs{\NVU}.

The question of the transplanckian modes and that of unitary evolution are both connected to another question, that of where radiation from a black hole originates. A common view has been that Hawking radiation originates in  high energy excitations that are very near the horizon, but physical tests, based for example on the Stefan-Boltzman law and on the behavior\refs{\DFU\ChFu-\CGHS} of the stress tensor, have suggested a different interpretation, in which these excitations originate in a ``quantum atmosphere" with depth comparable to the Schwarzschild radius\refs{\SGStBo} (see also \DLP, and \refs{\Unru\Full-\Bard} for earlier related arguments).  This question is important because it also helps guide understanding of where Hawking's analysis might be modified.  For example, if high-energy modes near the horizon did play an important role, and their evolution is also assumed to be modified, the result is a state that an infalling observer perceives to contain high energy excitations\refs{\SGTrieste\BPZ\AMPS-\AMPSS}.  Such a ``firewall" description was particularly advocated in \AMPS.
 
 If, on the other hand, the Hawking radiation originates in a broader vicinity of the horizon, say comparable to its size, and if Hawking's description is modified on these scales, that suggests a very different, ``nonviolent" picture\refs{\SGmodels\BHQIUE-\NVearly,\NVU} of the unitarization of the Hawking process.  
 
This paper will investigate the evolution of the Hawking state, in a Schr\"odinger picture treatment that is based on smooth slices that cut through the horizon and into the black hole interior.\foot{For a preliminary discussion of this approach, see \NVU; for a related discussion, see \refs{\BHS,\HoSi}.}   This achieves an improved description of the state, complementary to that of \Hawkrad.  It also addresses the transplanckian problem: as expected, the evolution does not exhibit a role for very high-energy, or short wavelength, excitations near the horizon.  By comparing with the original description given by Hawking, it is seen that transplanckian effects are an artifact of the mode basis that Hawking chose to analyze the radiation, and that basis of course becomes singular at the horizon.  This artifact is removed in a different, ``regular," basis.  

This regular description of the state then supports the arguments\refs{\SGStBo} that the Hawking radiation does indeed originate in excitations with horizon-size wavelengths, in a comparably-sized region of the horizon.  The description of the evolving quantum state of Hawking radiation then can serve as a background on which to study effects that unitarize the evolution, which are expected to be operative on macroscopic rather than microscopic scales, for a large black hole,
extending the work of \refs{\NVU}.  

In outline, the next section describes various time slicings of a Schwarzschild background, and describes general Schr\"odinger evolution on such slicings.  Section 
three turns to study Hawking evolution on smooth slicings that enter the horizon.  Very explicit examples can be provided for two-dimensional black holes.  Calculation of the quantum hamiltonian reveals a description of the creation of the Hawking excitations, and this occurs at longer wavelengths, rather than microscopic wavelengths.  Other aspects of the evolving state are also studied, including the pairing (entanglement) between excitations inside and outside the horizon.  The last section closes with further discussion of the transplanckian question, of generalizations to other black holes and to cosmology, and of a possible connection to tunneling calculations of Hawking radiation\refs{\PaWi}.  It also discusses the question of including backreaction and evolution of the black hole spacetime, and that of the ultimate unitarization of the evolution.

\newsec{Schr\"odinger evolution between slices}

\subsec{Slicing black hole spacetimes}

A dynamical desciption of the evolution of the quantum state in the vicinity of a black hole (BH) can be given by introducing a time-slicing for the BH spacetime.  We will focus on  Schwarzschild BHs, but the discussion should extend to more general BHs.  The $D$-dimensional Schwarzschild geometry is
\eqn\Schw{ds^2 = - f(r) dt^2 + {dr^2\over f(r)} + r^2 d\Omega^2_{D-2}\ ,}
with dimension-dependent function $f(r)=1-\mu(r)$.
For dimensions $D>3$,
\eqn\HDbh{\mu(r)=\left({R\over r}\right)^{D-3}\ ,}
where $R$ is the Schwarzschild radius.  Eq.~\Schw\ also extends to
two-dimensional BHs\refs{\WittBH,\CGHS}, with
\eqn\TDbh{\mu(r)=e^{-2(r-R)}\ .}
In general $\mu(R)=1$, and $\mu(r)$ diverges at the singularity, $r=0$ for $D>3$, and $r=-\infty$ for $D=2$.
It is also useful to consider the metric in ingoing Eddington-Finkelstein coordinates,
\eqn\EF{ds^2=-f(r) dx^{+2} + 2 dx^+ dr + r^2 d\Omega^2_{D-2}\ .}
to eliminate the coordinate singularity at the horizon.

The geometry \Schw\ does not fully describe a quantum BH, since BH radiation will decrease the mass $M$ of the BH.  However, the fractional change in  $M$ due to the emission of one quantum of typical energy $\sim 1/R$, over a time $\sim R$, is $\calo(1/MR)$, so a large quantum BH is expected to be approximately described over a period of many emissions by a stationary geometry of the form \Schw\ or \EF.  Stationarity corresponds to the symmetry $x^+\rightarrow x^+ + \epsilon$ in \EF.

 \Ifig{\Fig\slices}{Shown are the four types of slices described in the text, in an Eddington-Finkelstein diagram based on ingoing coordinates.  In addition to the familiar Schwarzschild slices, there are nice slices, which asymptote to a constant $r=R_n$,  natural slices, which reach $r=0$, and  straight slices, which are a special case of the latter.  All slices asymptote to Schwarzschild time slices as $r\rightarrow\infty$.  The family of slices used to parameterize the geometry is found by translating one of these slices vertically in the figure, which corresponds to a Schwarzschild time translation, $t\rightarrow t+\Delta t$.}{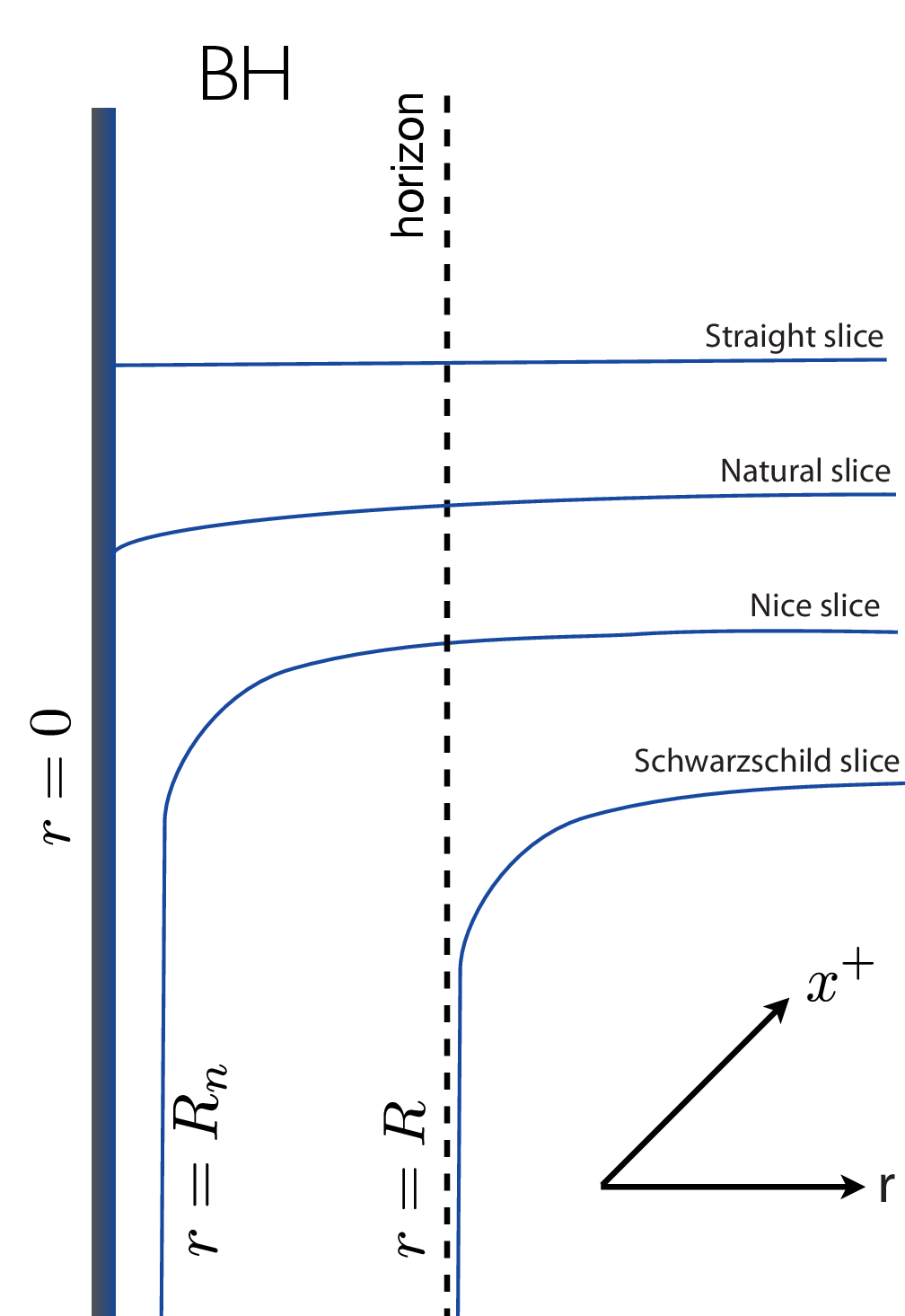}{2.8}

Spacelike Schwarzschild slices of constant $t$ remain outside the horizon.  Smooth transhorizon spatial slices\foot{Previous use of certain such slices to study Hawking radiation includes \refs{\JacoC\MeWeone\MeWetwo-\JacoRev}.} can be defined as in \refs{\BHQIUE,\NVU} by introducing a function $s(r)$ that asymptotes to $r$ as $r\rightarrow\infty$.  Then, for a given time parameter $T$, the slice is given by the solution to the equation 
\eqn\slicedef{T=x^+-s(r)\ .}
Asymptotically as $r\rightarrow\infty$, these slices match the constant $t$ slices with $t=T$, and under $T\rightarrow T+ \epsilon$, the slices translate by the Schwarzschild time-translation symmetry.  The configuration of the slice near and within the BH depends on the behavior of $s(r)$ there.  If $s(r)$ is finite at the singularity, the slices hit the singularity.  We refer to such slices as {\it natural slices}\refs{\NLvC}, since they exhibit the behavior corresponding to time evolution naturally pulling observers to $r=0$.  A very simple example is the case\refs{\NVU} $s(r)=r$ of ``straight" slices, and another arises from Painlev\'e-Gullstrand coordinates (see, {\it e.g.}, \refs{\MaPo}).  If $s(r)\rightarrow-\infty$ at some $r=R_n$ before reaching the singularity, the slices instead avoid the singularity.  Taking $R_n<R$ gives examples of {\it nice slices}\refs{\LPSTU\QBHB-\Mathur,\BHQIUE,\NVU}.  A special case of $R_n=R$ returns one to the Schwarzschild $t$ slices.  The different cases are illustrated in \slices.

We will study the evolution of the quantum states on such slices.  To do so, it is useful to put the metric \Schw, \EF\ in Arnowitt Deser Misner (ADM) form\refs{\ADM}, 
\eqn\admmet{ds^2= -N^2 dT^2 + q_{ij}(dx^i+N^i dT)(dx^j+N^j dT)\ .}
The lapse $N$, shift $N^i$, and spatial metric $q_{ij}$ for the slicing determined by $\slicedef$ are given by\NVU
\eqn\admvars{N^2={1\over s'(2-fs')}\quad ,\quad N_r={1-fs'}\quad ,\quad q_{rr}= s'(2-fs')\ ,}
where $N_i=q_{ij}N^j$ and $s'=ds/dr$.  A useful alternate choice of radial coordinate is to use
\eqn\rhodef{\rho=s(r)\ .}
Stationarity of \admmet\ is seen through its $T$ independence.  The unit normal to the slices is given by
\eqn\normdef{n^\mu=(1,-N^i)/N\ .}

\subsec{Schr\"odinger picture evolution in curved spacetime}

For simplicity we consider evolution of a massless scalar $\phi$, with lagrangian
\eqn\lagr{{\cal L} = - \hf g^{\mu\nu} \partial_\mu \phi \partial_\nu \phi\ ,}
but the general analysis extends to other fields.  Schr\"odinger evolution is based on the canonical evolution.  With the metric in ADM form \admmet, the canonical momentum is
\eqn\canonp{\pi= {\partial_T \phi - N^i \partial_i\phi\over N}= n^\mu\partial_\mu\phi\ ,}
and the hamiltonian becomes
\eqn\admham{H = \int d^{D-1} x \sqrt{q} \left[ \hf N(\pi^2 + q^{ij}\partial_i \phi \partial_j \phi )+ N^i \pi\partial_i\phi \right]\ .}
Quantization is described by introducing the canonical commutators
\eqn\CCR{[\pi(x^i,T),\phi(x^{i\prime},T)] = -i{\delta^{D-1}(x-x')\over \sqrt{q}}}
and the corresponding representation $\pi=-i\delta/\delta\phi$.  Then, the evolution operator is
\eqn\evolop{U(T_2,T_1) = \exp\left\{ -i\int_{T_1}^{T_2} HdT \right\}\ .}
For a time-dependent hamiltonian $H$, which would result from a time-dependent background metric \admmet, there are subtleties in defining the corresponding Schr\"odinger evolution\refs{\ToVa\Cortetal\AgAs-\MuOe}, but those are avoided for the stationary geometries \admmet, \admvars.

In order to give a Fock space representation for the states and their evolution, we need to introduce an appropriate basis of mode functions.  In general there is arbitrariness in this choice.  We can think of specifying such solutions to the equations of motion by giving the data $\gamma_i(x)= (\phi_i(x), \pi_i(x))$ at some time $T$.  To quantize, one needs a division into the analog of positive and negative frequency modes, which can be provided by giving a mode basis that has a complex structure\refs{\Kay,\More,\AgAs}\ $J$ distinguishing the modes,
\eqn\Jdef{J\gamma_A = i \gamma_A\quad,\quad J\gamma_A^* = -i\gamma_A^*\ .}
(Examples will be provided shortly.)  Then we use the expansions
\eqn\modeexp{\phi(x^i,T) = \sum_A\left[a_A\phi_A(x^i) + a_A^\dagger \phi_A^*(x^i)\right]\quad,\quad \pi(x^i,T) = \sum_A\left[a_A \pi_A(x^i) + a_A^\dagger \pi_A^*(x^i)\right]\ .}
 If the mode basis is normalized such that
\eqn\modenorm{(\gamma_A,\gamma_B) =  \delta_{AB}\quad,\quad (\gamma_A,\gamma_B^*)=0\ ,}
with the norm (inherited from the symplectic form)
\eqn\normdef{(\gamma_1,\gamma_2)=i\int d^{D-1} x \sqrt{q} (\phi_1^* \pi_2 - \pi_1^*\phi_2)}
then the operators $a_A,a_B^\dagger$ satisfy the commutators
\eqn\acomms{[a_A,a_B^\dagger]=\delta_{AB}\quad,\quad[a_A,a_B]=[a_A^\dagger,a_B^\dagger]=0\ .}

The state that corresponds to the vacuum in this choice of basis at an initial time $T=T_0$ is the state $|\psi,T_0\rangle=|0\rangle$
satisfying
\eqn\vacdef{a_A|0\rangle=0\ ,}
and excitations on this are built with the $a_A^\dagger$.
Schr\"odinger evolution of an initial state is then described by the  operator \evolop, which can be rewritten in terms of the ladder operators $a_A, a_A^\dagger$.  In general, the initial vacuum state evolves into a state that does not satisfy the vacuum condition \vacdef\ at a later time.  This contrasts with the usual Heisenberg picture evolution, in which the state is constant but the operators $\phi$ and $\pi$ evolve with time.

\newsec{Hawking evolution on smooth slices}

We next combine the preceding general description of Schwarzschild evolution with the slicings of the previous subsection, to describe dynamics in a BH background.
For simplicity, this paper will focus on the two-dimensional case, but this work can be extended to higher dimensions by using a spherical wave decomposition and the resulting 2d evolution with effective potentials.

\subsec{Mode bases}

Consider evolution on slices determined by a 
general choice of  $s(r)$.  A first step is to describe a suitable basis of modes.  A 2d massless scalar decomposes into separate left- and right-moving parts.  This is seen explicitly by rewriting the 2d metric \Schw, \TDbh\ as
\eqn\tdbh{ds^2=-{dX^+ dX^-\over M-X^+X^-}\ ,}
using\foot{We work in units where the parameter $\lambda$ of \refs{\CGHS} is set to one.}
\eqn\rtoX{e^{2r}= M-X^+X^-\ ,}
$X^\pm=\pm e^{\pm x^\pm}$, and $2t=x^++x^-$.  Here the horizon $r=R$ corresponds to $X^-=0$ or $X^+=0$, and $M=e^{2R}$; the right BH exterior is $X^+>0$, $X^-<0$.  
Left and right movers are then general functions of $X^+$ and $X^-$, respectively.  Since we are interested in radiation, we focus on right-moving modes.
In the slice coordinates $T,r$, the right-moving condition $\partial_+ \phi=0$ becomes
\eqn\RMTr{\partial_T\phi = -{f\over 2-s'f}\partial_r\phi\ ,}
or, with $\pi$ given by \canonp, 
\eqn\RMp{\pi = - {\partial_r\phi \over \sqrt{q_{rr}}}\ .}
There are two particularly natural choices for mode bases for these right movers.  

\subsubsec{Energy eigenmodes}

The first choice is to consider definite frequency modes with respect to $T$; in the exterior region, these are  $e^{-i\omega x^-}$.
These modes feature prominently in Hawking's original calculation\refs{\Hawkrad}.  Using the coordinate relation
\eqn\XtorT{X^-=-e^{-x^-}=-2\sinh(r-R) e^{R-T-s(r)+r}\ ,}
derived from \slicedef\ and \rtoX,
we see that these modes become singular at the horizon.  Similar modes are defined inside the horizon, $X^->0$, by using a new coordinate
$\hat x^-$, 
\eqn\xhatdef{X^-=e^{\hat x ^-}\ ,}
to define modes $e^{-i\omega \hat x^-}$.  These are also clearly singular at the horizon. 

On a given time $T$ slice, the outside modes correspond to the initial data (from \RMp)
\eqn\omps{\gamma_\omega(r)=( e^{-i\omega x^-(r)},{i\omega\over\sqrt{q_{rr}}} {\partial x^-\over \partial r} e^{-i\omega x^-(r)})\ }
and similarly for the inside modes.
The inner product \normdef\ then becomes
\eqn\omprod{(\gamma_{\omega},\gamma_{\omega'}) = 4\pi\omega\delta(\omega-\omega')\ ,}
and we can expand outside and inside solutions in the forms
\eqn\omexp{\phi(x) = \int {d\omega\over 4\pi\omega} \left[b_\omega e^{-i\omega x^-}+ b_\omega^\dagger e^{i\omega x^-} \right]\quad,\quad 
\phi(x) = \int {d\omega\over 4\pi\omega} \left[\hat b_\omega e^{-i\omega \hat x^-}+ \hat b_\omega^\dagger e^{i\omega \hat x^-} \right]\ ,}
 with normalizations
\eqn\omcomm{[b_\omega,b^\dagger_{\omega'}] = [\hat b_\omega,\hat b^\dagger_{\omega'}]=4\pi\omega\delta(\omega-\omega')\ .}
Corresponding to the singularity of these modes at the horizon, the vacuum $|0\rangle_B=|0\rangle|\hat 0\rangle$ annihilated by $b_\omega$ and $\hat b_\omega$ is also singular there; it is the 2d version of the Boulware vacuum\refs{\Boul}.  While states obtained by acting on this vacuum by the corresponding creation operators are eigenstates of the hamiltonian, they are consequently  not expected to be physical states.

\subsubsec{Regular modes}

Physical states can be more easily described by using modes that are regular at the horizon.  Useful examples are either the modes 
\eqn\rmodes{\phi_k=e^{ikr}\quad  {\rm or} \quad \phi_k=e^{ik\rho}\ ,}
on a constant $T$ slice, with the latter defined using the radial coordinate \rhodef.  The right-moving condition \RMp\ then implies corresponding momenta
\eqn\prmodes{\pi_k= -{ik\over \sqrt{q_{rr}}} e^{ikr}\quad {\rm or} \quad \pi_k= -{ik\over \sqrt{q_{\rho\rho}}} e^{ik\rho}\ ,}
respectively.  In either case the modes $\gamma_k=(\phi_k,\pi_k)$ satisfy the orthonormality condition
\eqn\normmodes{(\gamma_k,\gamma_{k'})=4\pi k\delta(k-k')\ ,}
as in \omprod. Both inside and outside the horizon, the field and momentum can be written
\eqn\kexp{\phi = \int_0^\infty {dk\over 4\pi k}\left[ a_k \phi_k + a_k^\dagger \phi_k^*\right]\quad,\quad \pi = \int_0^\infty {dk\over 4\pi k}\left[ a_k \pi_k + a_k^\dagger \pi_k^*\right]\ ,}
with commutators
\eqn\acomms{[a_k,a_{k'}^\dagger] = 4\pi k\delta(k-k')\ .}
The vacuum $|0\rangle$, satisfying
\eqn\rvac{a_k|0\rangle=0\ ,}
is now regular at the horizon, but since it is not an energy eigenstate, has non-trivial evolution.

An important feature of these regular modes is that they match the energy eigenmodes increasingly well as $r\rightarrow\infty$.  This can be seen directly from the coordinate transformation \XtorT, and the fact that $s(r)\rightarrow r$ in this limit.\foot{They also closely match the modes $e^{-ikX^-}$ near the horizon, where the latter are also regular, but these $X^-$ modes are not well-behaved at infinity.}

\subsec{Quantum hamiltonian}

Evolution in either basis of modes is determined by the quantum version of the hamiltonian, found by inserting the mode expansions \omexp\ or \kexp\  into \admham.  For right-movers, using the sliced metric \admvars\ and the right-moving condition \RMp, the ADM hamiltonian \admham\ simplifies to
\eqn\hamr{H=\int dr {f\over 2-f s'} (\partial_r\phi)^2\ .}

For the energy eigenmodes \omps, this hamiltonian becomes
\eqn\hamen{H=\hf \int {d\omega\over 4\pi\omega} \, \omega \left(b^\dagger_\omega b_\omega - \hat b^\dagger_\omega \hat b_\omega + h.c.\right) = \int {d\omega\over 4\pi\omega}\, \omega \left(b^\dagger_\omega b_\omega - \hat b^\dagger_\omega \hat b_\omega \right)\ ,}
as anticipated; note that the normal-ordering constant cancels.  Notice also that this expression exhibits negative energies for the modes inside the horizon, in line with  common statements. For the regular modes, we find the more complicated expression
\eqn\hamreg{\eqalign{H&=\int {dk\over 4\pi} {dk'\over 4\pi} \left[ A(k,k') a_k^\dagger a_{k'} + B(k,k') a^\dagger_k a^\dagger_{k'} + h.c.\right]\cr  &= \int {dk\over 4\pi} {dk'\over 4\pi} \left[ 2 A(k,k') a_k^\dagger a_{k'} + B(k,k') a^\dagger_k a^\dagger_{k'} + B^*(k,k') a_k a_{k'}\right]+E_0\ .}}
with coefficients
\eqn\ABdefs{ A(k,k')= \int_{-\infty}^\infty dr {f\over 2-fs'} {\partial_r \phi_k^* \partial_r\phi_{k'}\over kk'} \quad,\quad B(k,k')= \int_{-\infty}^\infty dr {f\over 2-fs'}{\partial_r \phi_k^* \partial_r\phi_{k'}^*\over kk'}\ ,}
and with $E_0$ a normal-ordering constant which can be subtracted from $H$.  The form of the hamiltonian \hamreg\ makes the non-trivial evolution in the regular bases clear; it includes creation of excitations through time evolution, in these bases.

\subsec{Evolution and structure of the state}

We now consider evolution of a state that is the vacuum $|0\rangle$ of \rvac\ at an initial time, which may be chosen to be $T=0$, so
\eqn\stateevol{|\psi,T\rangle = e^{-i H T}|0\rangle\ .}
 This state is regular at the horizon, and is a good candidate for the initial state of the BH.  Other states that are regular at the horizon can include additional initial excitations in its vicinity, but those excitations in general either evolve to infinity or to the singularity in a short time; the condition of regularity at the horizon is expected to be the key condition governing the correct long-time evolution.

The long-time evolution of \stateevol\ will then yield continual production of excitations from the vicinity of $r=R$, as seen from the hamiltonian \hamreg.  At $r=\infty$, as noted above, these behave just like ordinary positive-energy flat space excitations; they comprise the Hawking emission from the BH.

There are various checks on this.  For example, while the full evolution of the regular modes via $H$ of \hamreg\ is somewhat complicated,
the high-energy spectrum of the Hawking radiation can be inferred directly from this expression.  Consider, for simplicity, the case of straight slices, $s(r)=r$.  In this case we find
\eqn\Bstraight{B(k,k')=- e^{-i(k+k')R} \int_{-\infty}^\infty dx \tanh x\, e^{-i(k+k')x} =  {i\pi \over \sinh[\pi(k+k')/2]}e^{-i(k+k')R}\ .}
At large $k$, $k'$, $B(k,k')\propto e^{-\pi( k+k')/2}$; for $k=k'$ this  gives the expected thermal spectrum with temperature\foot{See, {\it e.g.}, \refs{\GiNe}.} $T=1/(2\pi)$.  Notice also  from \Bstraight\ that the dominant excitations are produced at wavelengths with $k=\calo(1)$, and so there is no direct role for very high energy modes; these are only produced with the corresponding exponential suppression. 

Another aspect of the Hawking state is the characteristic correlation between outgoing excitations and BH modes; their entanglement is central to the information problem/unitarity crisis.  These correlations are of course present in the description in terms of regular modes, but are more obscure.  One sees directly that the hamiltonian \hamreg\ creates correlated pairs of $k$ excitations, and the phase in \Bstraight\ indicates that they are created in the vicinity of $r=R$.  
One way of characterizing the state $|\psi,T\rangle$ is to rewrite the condition $a_k|0\rangle=0$ as
\eqn\statecond{e^{-iHT} a_k e^{iHT} |\psi,T\rangle =0\ ,}
and the correlation between excitations can for example be seen by expanding \statecond\ in $H$.  While in this basis it is more difficult to exhibit the trans-horizon nature of the correlations, one finds an initial indication of this by examining the time evolution of the $T=0$ mode $\tilde \phi_k=e^{ik(r-R)}$, which also simplifies in the straight slicing to the form, from \XtorT,
\eqn\straightevol{\tilde\phi_k(T,r)= e^{ ik\sinh^{-1}\left[e^{-T} \sinh (r-R)\right]}\ .}
This solution has an outgoing part with $r\sim T+R$, and an ingoing part with $r\sim R-T$.

The trans-horizon correlations are most easily seen by using the energy eigenbasis, and the equality of the expressions \hamen\ and \hamreg\ for the hamiltonian.  For example, approaching infinity, as we have seen, the outgoing modes in the two bases become identical.  Moreover, the time translation symmetry ensures conservation of $H$.  From the expression \hamen, this conservation shows that an outgoing positive energy excitation should be paired with an effectively negative energy excitation, created by the internal operators $\hat b_\omega$.  

This pairing can be made more explicit, in an argument extending  \refs{\UnruNotes,\HawkUnc,\GiNe}.  First, we have noted that the transform to exterior coordinate $x^-$ given by \XtorT\ is singular at $r=R$,
\eqn\xmxm{x^- = -\ln\left[2\sinh(r-R)\right] + T-R+s(r)-r\ .}
However, if we analytically continue in the complex $r$ plane by placing the branch cut in the lower half plane, the resulting analytic continuation $f^-(r)$ is analytic in the upper half $r$ plane (assuming analyticity for $s(r)$), and gives
\eqn\xmcont{f^-=\cases{ x^- &if $r>R$\cr - i\pi -\ln\left[2\sinh(R-r)\right] + T-R+s(r)-r=- i\pi-\hat x^- &if $r<R$}\ . }
%
Therefore, the functions 
\eqn\posfreq{e^{-i\omega f^-(r)}= \theta(r-R) e^{-i\omega x^-} + \theta(R-r)e^{-\pi\omega} e^{i\omega \hat x^-}}
are positive frequency with respect to $r$, and the corresponding operators
\eqn\anncomb{c_\omega=\zeta_\omega(b_\omega - e^{-\pi\omega} \hat b^\dagger_\omega)\ ,}
with $\zeta_\omega = 1/\sqrt{1-e^{-2\pi\omega}}$, 
annihilate the state $|\psi,0\rangle=|0\rangle$.  Likewise, we can define the analytic continuation $\hat f^-(r)$ of $\hat x^-$
that places the branch cut in the lower half $r$ plane and is analytic in the upper half plane, giving
\eqn\xmconth{\hat f^-=\cases{ \hat x^- &if $r<R$\cr - i\pi +\ln\left[2\sinh(r-R)\right] - T+R-s(r)+r=- i\pi- x^- &if $r>R$}\ . }
So, the functions
\eqn\posfreqh{e^{-i\omega \hat f^-(r)} = \theta(R-r) e^{-i\omega \hat x^-} +  \theta(r-R) e^{-\pi\omega}e^{+i\omega x^-}}
are also positive frequency with respect to $r$, and the corresponding operators
\eqn\anncombh{\hat c_\omega=\zeta_\omega(\hat b_\omega - e^{-\pi\omega} b^\dagger_\omega)\ }
also annihilate $|\psi,0\rangle$.  

The equations
\eqn\pairing{0=(c^\dagger_\omega c_\omega - \hat c^\dagger_\omega \hat c_\omega)|\psi,0\rangle = (b^\dagger_\omega b_\omega - \hat b^\dagger_\omega \hat b_\omega)|\psi,0\rangle\ ,}
for each $\omega$, then imply the precise pairing and entanglement between inside and outside excitations.  Schematically, 
the equality of the number operators in \pairing\ implies
\eqn\pairstate{|\psi,0\rangle\sim \sum_{\{n_\omega\}} C\left(\{n_\omega\}\right) |\widehat{\{n_\omega\}}\rangle |\{n_\omega\}\rangle\ ,}
where $ |\{n_\omega\}\rangle$, $ |\widehat{\{n_\omega\}}\rangle$ are occupation number eigenstates, and $C\left(\{n_\omega\}\right)$ are coefficients   that can be determined by the conditions that $c_\omega$ and $\hat c_\omega$ annihilate the state.  The result is 
\eqn\pairstateb{|\psi,T\rangle \sim C \sum_{\{n_\omega\}} e^{-\pi \int d\omega  \omega n_\omega} |\widehat{\{n_\omega\}}\rangle |\{n_\omega\}\rangle\ ,}
with $C$ a normalization constant.
However, this expression is somewhat formal, since the state $|\psi,T\rangle$ does not actually lie in the product Hilbert space, due to the type-III property of the von Neumann algebra ({\it i.e.} infinite entanglement). 
The pairing conditions for  $|\psi,T\rangle$ can alternately be formulated in terms of the $a_k$ operators, by transforming back to that basis.  This pairing condition also ensures regular behavior of interactions with infalling matter.  This looks unexpected from the viewpoint of the energy eigenbasis, since the state contains high energy $b^\dagger_\omega$ and $\hat b^\dagger_\omega$ excitations near the horizon, but their pairing leads to a cancellation between the corresponding pieces of the stress tensor\refs{\BHIUN} and its contribution to gravitational interactions.

Another aspect of the state that is evident from \hamr\ is the freezing of the evolution\refs{\QBHB} at $r=R_n$ in the case of a nice slicing, due to infinite $s'$ forcing the hamiltonian density to vanish.  Other features of the state can be investigated, but are deferred for future work, where modifications\refs{\NVU} that unitarize evolution will also be further investigated.

\newsec{Consequences, connections, and generalizations}

Hawking's original derivation \refs{\Hawkrad}, and subsequent rederivations, have used methods which exhibit a transplanckian problem, of referring to ultrahigh energy excitations near the horizon.  In the present description, that corresponds to formulating the description of the state in terms of the energy eigenmodes, which are singular at the horizon precisely due to this transplanckian behavior.  This has led to a lot of discussion of the role and meaning of transplanckian modes, suspicion that the calculation should be modified, {\it etc.}  

However, the preceding analysis should help make it clear that these problems are an artifact of using a singular basis for the modes.  If one alternately uses a regular basis, like that of the regular modes, we have seen that there is no explicit role for transplanckian excitations.  The price paid is that the evolution, governed by \hamreg, is more complicated to describe. 
The relevant dimensionful scales are of order the Hawking temperature $T=1/(2\pi)$ and the state evolves adiabatically on shorter length scales.  Indeed, it appears possible (though we will not explicitly do so here) to formulate the description of the evolving state  $|\psi,T\rangle$ in the context of a cutoff version of the theory, with cutoff at a subplanckian $k$, eliminating any transplanckian dependence.\foot{Such a procedure connects with a proposal\JacoC\ to implement a cutoff based on frequencies measured by infalling observers.}  

While for simplicity the preceding analysis focussed on the two-dimensional case, the analysis should generalize to the higher-dimensional case.  Then, the scales entering the expressions \hamreg, \ABdefs, and \Bstraight\ will be set by the horizon radius $R$.  

The role of the ultrahigh-energy modes in Hawking's calculation has also led some to conclude that the Hawking radiation is ``produced" in a region microscopically close to the horizon.  In contrast, arguments have been given based on physical tests that it is more physical to describe the Hawking radiation as produced in a ``quantum atmosphere" region of size $\Delta r\sim R$ near the horizon\refs{\SGStBo} (see also \refs{\DLP}).  The description of the evolving state given above also supports the latter interpretation:  the Hawking modes, which escape to infinity with characteristic wavelength $\lambda\sim R$, are produced by the hamiltonian \hamreg\ in a region of size $\sim R$ near the horizon; this is directly connected to the absence of a role for ultraplanckian modes, which are also those ultranear the horizon.

The analogy between Hawking radiation and cosmological production of fluctuations during inflation is well known, and we expect that similar conclusions extend to that case as well: evolution can be described without invoking transplanckian excitations, and fluctuations are produced at a characteristic length scale given by the horizon size.  However, the non-trivial time dependence there is an additional complication which we defer to future work.

It is likewise expected that a similar analysis can be carried out for the cases of charged and rotating black holes.

It also appears that it should be possible to connect the analysis we have described, based on a stationary slicing of the metric, to efforts to provide a 
tunneling description of the Hawking process.  In particular, the analysis of \refs{\PaWi} is based on the Painlev\'e coordinates, which give a special case of a stationary slicing.  It seems plausible that a more general analysis, connecting to a first quantized picture, might be given.

Of course, the most important questions involve departures from the stationary evolution of the Hawking state.  Specifically, we can work perturbatively in the gravitational coupling $G$.  The outgoing particles will be gravitationally dressed\refs{\DoGione}, and will carry away energy and lower the BH mass. It is plausible that a perturbative description of the effect of this backreaction can also be treated, beginning with the evolving state description that we have given, by including the metric perturbations and their coupling to the stress tensor.

Once backreaction is accounted for, we know that we will encounter the ultimate problem of unitarity.  The preceding analysis reproduces the known entanglement of the Hawking radiation with internal excitations of the BH.  Local quantum field theory does not provide a mechanism for this entanglement to transfer to the outgoing radiation, so if the BH shrinks and disappears, unitarity is violated.  Thus, in order to save unitarity, new interactions appear to be required to transfer this entanglement to the outgoing state\refs{\NLvC,\SGmodels\BHQIUE-\NVearly,\NVU}.  If one assumes that these interactions are only present to modes within a microscopic cutoff scale of the horizon, a firewall results\refs{\SGTrieste\BPZ\AMPS-\AMPSS}.  While such a hypothesis is motivated by a picture in which the Hawking radiation is produced at these microscopic scales, the contrasting view, advocated in \SGStBo\ and supported by the analysis here, that Hawking excitations are produced in  a vicinity of size $\sim R$ near the horizon, also suggests that the new interactions that unitarize evolution are operative on these scales\refs{\SGmodels\BHQIUE-\NVearly,\NVU}.

It has been argued that such unitarizing interactions might be viewed as a small, in an appropriate sense, correction to the Hawking evolution.  If so, the analysis of this paper is also helpful for that, as it provides the background evolution of the state, on top of which the additional corrections of the interactions can be described\refs{\NVU}.  Specifically, starting with a parameterization of  Schr\"odinger evolution, it should be possible to describe these interactions in an effective approach in terms of additional contributions to the hamiltonian, which transfer entanglement from the BH state to outgoing radiation.  The explicit parameterization of the background Hawking state and evolution should serve as a useful tool in this analysis.

\bigskip\bigskip\centerline{{\bf Acknowledgments}}\nobreak

This material is based upon work supported in part by the U.S. Department of Energy, Office of Science, under Award Number {DE-SC}0011702. I thank G. Horowitz and D. Marolf for comments on a draft of this paper, and T. Jacobson for comments on its first version.

\listrefs
\end